\documentclass[letterpaper,notitlepage,times,12pt]{article}
\usepackage{graphicx}
\usepackage{rotating}
\usepackage{amssymb}
\usepackage{cite}
\usepackage{subfigure}
\usepackage{geometry}
\begin{document}

\date{October 2006}

\title{{\Large A VoIP Privacy Mechanism and its Application in VoIP Peering for Voice Service Provider Topology and Identity Hiding}}
\author{Charles Shen and Henning Schulzrinne \\
Department of Computer Science \\
Columbia University \\
\{charles, hgs\}@cs.columbia.edu
}

\maketitle

\abstract{Voice Service Providers (VSPs) participating in VoIP peering frequently want to withhold their identity and related privacy-sensitive information from other parties during the VoIP communication. A number of existing documents on VoIP privacy exist, but most of them focus on end user privacy. By summarizing and extending existing work, we present a unified privacy mechanism for both VoIP users and service providers. We also show a case study on how VSPs can use this mechanism for identity and topology hiding in VoIP peering.}

\section{Introduction}

Privacy mechanisms for SIP-based VoIP can be found in a number of existing documents, including current RFCs (RFC 3323~\cite{rfc3323}, RFC 3325~\cite{rfc3325}, RFC 3261~\cite{rfc3261}, RFC 3711~\cite{rfc3711}), an obsolete RFC (RFC 2543~\cite{rfc2543}), and an expired internet draft (Byerly\cite{siphideroute}). One motivation of our work is to summarize useful pieces of related contents in these documents and present them as a unified privacy mechanism for VoIP.

The other motivation of our work comes from requirements raised by Voice Service Providers (VSPs) that participate in VoIP peering. VoIP peering is the interconnection of VSPs at the session layer for exchanging SIP-based signaling among each other. The peering VSPs together form a large end-to-end SIP-enabled signaling network. This network allows individual peering VSPs to greatly extend the service diversity and coverage they can provide to their customers, bypassing the Public Switched Telephone Network (PSTN). The ability to avoid routing traffic through PSTN also cuts significant service costs. Despite the advantages of peering, many VSPs concern about their identity and network topology information being exposed to other service providers in the peering network. VSP identity exposure potentially enables the VSP's competitors to learn about who its customers are and then target specific marketing efforts to win them over. VSP topology information exposure could reveal the VSP's internal network structure, network size and raise business concerns as well as security threats. Therefore, VSPs frequently desire to protect their privacy-sensitive information in VoIP peering. Existing privacy work on VoIP, however, mostly focuses on privacy requirements for end users. For example, RFC 3323 defines privacy as ``the withholding of the identity of a person (and related personal information) from one or more parties in an exchange of communications, specifically a SIP dialog. These parties potentially include the intended destination(s) of messages and/or any intermediaries handling these messages.'' Although most user privacy mechanisms also provide certain privacy protection for service providers, directly applying user privacy mechanisms does not solve the service provider privacy problem completely. Moreover, protecting service provider privacy does not always entail protecting user privacy at the same time.

In this report, we broaden the RFC 3323 definition of privacy to cover service providers. Specifically, privacy in this document is defined as ``the withholding of the identity of a person (and related personal information) and/or a service provider (and related service provider information, such as its network topology) from one or more parties in an exchange of communications, specifically a SIP-based VoIP session.'' We extend existing work and present a mechanism for both user and service provider privacy. We start with identifying message fields in key VoIP protocols that potentially reveal user or service provider information. Then we describe the privacy service functions that can be performed at the user side and at the network intermediaries. We show how user level and service provider level privacy requirements can be supported by those privacy functions. Using the service provider privacy component of the privacy mechanism, we gave a case study on how the VSPs can achieve identity and topology hiding in a generic VoIP peering architecture.

The rest of this report is organized as follows: Section~\ref{sec:voipprivacy} describes a privacy mechanism for SIP-based VoIP; Section~\ref{sec:pps} introduces VoIP peering architecture and gives a case study of achieving VSP privacy in VoIP peering. Section~\ref{sec:conclusion} concludes the report.

\section{A Privacy Mechanism for SIP-based VoIP} \label{sec:voipprivacy}

\subsection{Privacy-Sensitive Message Fields in SIP, SDP, RTP and RTCP} \label{sec:privacyfield}

The SIP header fields that may reveal user or service provider information are summarized in Table~\ref{tab:siphdreg}. The first column gives the header type with an example header format, some of which are drawn from Section 20 of RFC 3261. The second column indicates the privacy characteristics of the corresponding field. Since our privacy definition includes privacy-sensitive information for both user and service provider, this column contains a ``u'' for user-related information fields, and a ``p'' for service provider-related information fields.

\begin{table}[!th]
\begin{center}
\begin{tabular}{|l|c|} \hline
Header field example & privacy \\
\hline\hline
{\sf Alert-Info: \textless http://vsp.example.com/sounds/moo.wav\textgreater }& p\\ \hline
{\sf Authorization: Digest username=``Alice'', realm=``VIP@example.com'', }& u \\
{\sf \hspace{10pt}nonce=``84a4cc6f3082121f32b42a2187831a9e'',}& \\
{\sf \hspace{10pt}response=``7587245234b3434cc3412213e5f113a5432''} & \\ \hline
{\sf Call-ID:f81d4fae-7dec-11d0-a765-00a0c91e6bf6@192.0.2.4} & up \\ \hline
{\sf Call-Info: \textless http://vsp.example.com/alice/photo.jpg\textgreater;purpose=icon,} & p \\
{\sf \hspace{10pt}\textless http://vsp.example.com/alice/\textgreater;purpose=info} & p \\ \hline
{\sf Contact: ``Alice'' \textless sips:alice@vsp.example.com\textgreater;expires=60} & up \\ \hline
{\sf Error-Info: \textless sip:not-in-service-recording@vsp.example.com\textgreater} & p \\ \hline
{\sf From: ``Bob'' \textless sip:bob@vspexample.com\textgreater;tag=hyh8} & up \\ \hline
{\sf In-Reply-To: 10230@vsp.example.com,} & up \\
{\sf \hspace{10pt} 44150@vsp.example.com} & \\ \hline
{\sf Organization: example Inc.} & up \\ \hline
{\sf Proxy-Authenticate: Digest realm=``VIP@example.com'',} & p \\
{\sf \hspace{10pt} domain=``sip:vsp.example.com'', qop=``auth'',} & \\
{\sf \hspace{10pt} nonce=``f84f1cec41e6cbe5aea9c8e88d359'',} & \\
{\sf \hspace{10pt} opaque=``'', stale=FALSE, algorithm=MD5} & \\ \hline
{\sf Record-Route: \textless sip:vsp1.example.com;lr\textgreater,} & p \\
{\sf \hspace{10pt} \textless sip:vsp2.example.com;lr\textgreater} & \\ \hline
{\sf Reply-To: Bob \textless sip:bob@vsp.example.com\textgreater} & up \\ \hline
{\sf Route: \textless sip:vsp1.example.com;lr\textgreater,} & p \\
{\sf \hspace{10pt} \textless sip:vsp2.example.com;lr\textgreater} & \\ \hline
{\sf Server: HomeServer v2} & p \\ \hline
{\sf Subject: Tech Support} & u \\ \hline
{\sf To: sip:+12125551212@vsp.example.com} & up \\ \hline
{\sf User-Agent: Softphone Beta1.5} & p \\ \hline
{\sf Via: SIP/2.0/UDP vsp.example.com:5060;} & p \\
{\sf \hspace{10pt} branch=z9hG4bK87asdks7} & \\ \hline
{\sf Warning: 307 example.com ``Session parameter `foo' not understood''} & p \\ \hline
{\sf WWW-Authenticate: Digest realm=``VIP@example.com'',} & p \\
{\sf \hspace{10pt} domain=``sip:vsp.example.com'', qop=``auth'',} & \\
{\sf \hspace{10pt} nonce=``f84f1cec41e6cbe5aea9c8e88d359'',} & \\
{\sf \hspace{10pt} opaque=``'', stale=FALSE, algorithm=MD5} & \\ \hline
\end{tabular}
\end{center}
\caption{SIP header fields that contain privacy-sensitive user and service provider information}\label{tab:siphdreg}
\end{table}

Common user-related information includes the user name, email, SIP address, phone, location, and information about the subject of the session. Note that the host name or IP address where the session media starts is also considered user related information.

Service provider-related information is usually presented in the form of a domain name or IP address. This can be seen in {\sf Alert-Info}, {\sf Call-ID}, {\sf Call-Info}, {\sf Contact}, {\sf Error-Info}, {\sf From}, {\sf In-Reply-To}, {\sf Proxy-Authenticate}, {\sf Record-Route}, {\sf Reply-To}, {\sf Route}, {\sf To}, {\sf Via}, {\sf Warning}, and {\sf WWW-Authenticate} header fields.
Most of the header fields that reveal service provider information contain the user name or user display name and therefore also disclose user information. These header fields include {\sf Contact}, {\sf From}, {\sf In-Reply-To}, {\sf Reply-To}, and {\sf To}. In fact, all these header fields may contain SIP URI, which is commonly formed by a user part and a service provider part. SIP also allows the use of other URI formats. For example, tel URI~\cite{rfc3966} and Globally Routable User Agent URIs (GRUU)~\cite{gruu}. Some of the tel URI parameters, such as {\sf Phone-Context}, may expose identity about the service provider. The GRUU may contain a SIP address of record, and therefore may expose both user and service provider information.

The privacy characteristics of some of the header fields is less straightforward and depends highly on how they are used. For example, the {\sf Organization} header field contains the ``name of the organization to which the SIP element issuing the request or response belongs\cite{rfc3261}'', it may or may not be related to the user or the service provider; the {\sf Subject} header field ``provides a summary or indicates the nature of the call\cite{rfc3261}'' and will likely affect user privacy; the {\sf User-Agent} and {\sf Server} header fields contain information about the SIP user agent client and user agent server. In some cases, these information can be linked to the service provider, especially when the service provider is also providing the SIP user equipment.

SIP uses SDP to describe media streams in VoIP sessions. The SDP fields that may reveal user or service provider information are listed in Table~\ref{tab:sdpprivacy}. SIP session media is commonly carried in RTP\cite{rfc3550}. RTP is usually used together with its control protocol RTCP. One type of RTCP packets called Source Description (SDES) packets contain privacy-sensitive information as listed in Table~\ref{tab:rtpprivacy}. Most of the fields that reveal service provider information contain a domain name or IP address, similar to the case in SIP. The RTCP SDES {\sf tool} field is analogous to the SIP {\sf User-Agent} and {\sf Server} header fields. Whether it reveals information about the service provider depends on actual usage situation.

\begin{table}[!th]
\begin{center}
\begin{tabular}{|l|l|c|} \hline
Field name & Example & Privacy \\
\hline\hline
{\sf Origin} & {\sf o=charles 2890844526 2890842807 IN IP4 128.59.66.4} & up \\ \hline
{\sf Email} & {\sf e=charles@cs.columbia.edu} & up \\ \hline
{\sf Phone} & {\sf p=+12125551234} & u \\ \hline
{\sf Session-name} & {\sf s=VoIP seminar} & u \\ \hline
{\sf Information} & {\sf i=A seminar on VoIP} & u \\ \hline
{\sf URI} & {\sf u=http://www.cs.columbia.edu/~charles/sdp.ps} & up \\ \hline
{\sf Connection} & {\sf c=IN IP4 128.59.66.4} & up \\
\hline
\end{tabular}
\end{center}
\caption{SDP message fields related to user and service provider privacy}\label{tab:sdpprivacy}
\end{table}

\begin{table}[!th]
\begin{center}
\begin{tabular}{|l|l|c|} \hline
Field name & Example & Privacy \\
\hline\hline
{\sf Cname} & {\sf doe@sleepy.example.com} & up \\ \hline
{\sf Name} & {\sf Charles} & u \\ \hline
{\sf Loc} & {\sf Morningside, Manhattan} & u \\ \hline
{\sf Email} & {\sf charles@example.com} & up \\ \hline
{\sf Tool} & {\sf videotool 1.2} & p \\
\hline
\end{tabular}
\end{center}
\caption{RTCP SDES packet fields related to user and service provider privacy}\label{tab:rtpprivacy}
\end{table}

There are two important points we want to emphasize about the above privacy-sensitive fields. First, the ``u'' or ``p'' privacy characteristics of a particular field is assigned based on a common usage pattern. In some cases, there are different ways to fill in the field that can prevent it from exposing privacy-sensitive information, at the same time without affecting its validity. Indeed this property is used in the privacy mechanism discussed in the later part of this document. Second, the service provider information fields may actually correspond to different types of service providers. For example, the IP address in the SDP {\sf Origin} field may disclose the user's Internet Access Provider (IAP) identity; the domain name part of the SIP URI in the SIP {\sf Contact} header may disclose the user's VSP; the domain name in the RTCP SDES packet's {\sf Email} field may disclose the user's Email service provider. These different types of service providers could all be the same one or be completely different, or be of any other combination. Therefore, it is impossible to precisely correlate these fields to one particular type of service provider without knowing the actual service provider relationship.

\begin{table}[!h]
\begin{center}
\begin{tabular}{|l|c|c|c|c|c|c|c|c|} \hline
Header field & where & proxy & ACK & BYE & CAN & INV & OPT & REG \\
\hline\hline
{\sf Alert-Info} & R & ar & - & - & - & o & - & - \\
{\sf Alert-Info} & 180 & ar & - & - & - & o & - & - \\ \hline
{\sf Authorization} & R & & o & o & o & o & o & o \\ \hline
{\sf Call-ID} & c & r & m & m & m & m & m & m \\ \hline
{\sf Call-Info} & & ar & - & - & - & o & o & o \\ \hline
{\sf Contact} & R & & o & - & - & m & o & o \\
{\sf Contact} & 1xx & & - & - & - & o & - & - \\
{\sf Contact} & 2xx & & - & - & - & m & o & o \\
{\sf Contact} & 3xx & d & - & o & - & o & o & o \\
{\sf Contact} & 485 & & - & o & - & o & o & o \\ \hline
{\sf Error-Info} & 300-699 & a & - & o & o & o & o & o \\ \hline
{\sf From} & c & r & m & m & m & m & m & m \\ \hline
{\sf In-Reply-To} & R & & - & - & - & o & - & - \\ \hline
{\sf Organization} & & ar & - & - & - & o & o & o \\ \hline
{\sf Proxy-Authenticate} & 407 & ar & - & m & - & m & m & m \\
{\sf Proxy-Authenticate} & 401 & ar & - & o & o & o & o & o \\
{\sf Proxy-Authorization} & R & dr & o & o & - & o & o & o \\ \hline
{\sf Record-Route} & R & ar & o & o & o & o & o & - \\
{\sf Record-Route} & 2xx, 18x & mr & - & o & o & o & o & - \\ \hline
{\sf Reply-To} & & & - & - & - & o & - & - \\ \hline
{\sf Route} & R & adr & c & c & c & c & c & c \\ \hline
{\sf Server} & r & & - & o & o & o & o & o \\ \hline
{\sf Subject} & R & & - & - & - & o & - & - \\ \hline
{\sf To} & c(1) & r & m & m & m & m & m & m \\ \hline
{\sf User-Agent} & & & o & o & o & o & o & o \\ \hline
{\sf Via} & R & amr & m & m & m & m & m & m \\
{\sf Via} & rc & dr & m & m & m & m & m & m \\ \hline
{\sf Warning} & & r & - & o & o & o & o & o \\ \hline
{\sf WWW-Authenticate} & 401 & ar & - & m & - & m & m & m \\
{\sf WWW-Authenticate} & 407 & ar & - & o & - & o & o & o \\ \hline
\end{tabular}
\end{center}
\caption{Summary of SIP header fields}\label{tab:sumhdr}
\end{table}

\subsection{User Side Privacy Service Functions} \label{sec:ups}

This section discusses possible privacy functions the user side can perform.
For SIP and SDP, the user side functions are carried out by a SIP user agent. For RTP, the user side functions are carried out by an RTP client.
For convenience, we copied part of the table on ``Summary of (SIP) header fields'' from RFC 3261 in Table~\ref{tab:sumhdr}. We only list those header fields that could reveal privacy-sensitive information as indicated in Table~\ref{tab:siphdreg}. The notations used in Table~\ref{tab:sumhdr} are also copied as below:

\begin{quote}
The ``where'' column describes the request and response types in which the header field can be used.  Values in this column are:
\begin{quote}
\begin{description}
\item R: header field may only appear in requests;
\item r: header field may only appear in responses;
\item 2xx, 4xx, etc.: A numerical value or range indicates response codes with which the header field can be used;
\item c: header field is copied from the request to the response.
\end{description}
\end{quote}

An empty entry in the ``where'' column indicates that the header field may be present in all requests and responses.

The ``proxy'' column describes the operations a proxy may perform on a header field:

\begin{quote}
\begin{description}
\item a: A proxy can add or concatenate the header field if not present.
\item m: A proxy can modify an existing header field value.
\item d: A proxy can delete a header field value.
\item r: A proxy must be able to read the header field, and thus this header field cannot be encrypted.
\end{description}
\end{quote}

The next six columns relate to the presence of a header field in a method:

\begin{quote}
\begin{description}
\item c: Conditional; requirements on the header field depend on the context of the message.
\item m: The header field is mandatory.
\item m*: The header field SHOULD be sent, but clients/servers need to be prepared to receive messages without that header field.
\item o: The header field is optional.
\item t: The header field SHOULD be sent, but clients/servers need to be prepared to receive messages without that header field.  If a stream-based protocol (such as TCP) is used as a transport, then the header field MUST be sent.
\item *: The header field is required if the message body is not empty. See Sections 20.14, 20.15 and 7.4 for details.
\item -: The header field is not applicable.
\end{description}
\end{quote}

``Optional'' means that an element MAY include the header field in a request or response, and a user agent MAY ignore the header field if present in the request or response (The exception to this rule is the Require header field discussed in 20.32).  A ``mandatory'' header field MUST be present in a request, and MUST be understood by the user agent server receiving the request.  A mandatory response header field MUST be present in the response, and the header field MUST be understood by the user agent client processing the response.  ``Not applicable'' means that the header field MUST NOT be present in a request.  If one is placed in a request by mistake, it MUST be ignored by the user agent server receiving the request.  Similarly, a header field labeled ``not applicable'' for a response means that the user agent server MUST NOT place the header field in the response, and the user agent client MUST ignore the header field in the response.

\end{quote}

The user side can use two common methods to assist in privacy protection. The first method is removal and anonymization of privacy-sensitive information in the generated messages. For SIP message headers, the removal applies to SIP header fields that are optional in the signaling process. From Table~\ref{tab:sumhdr}, we see that the SIP {\sf Call-Info}, {\sf In-Reply-To}, {\sf Organization}, {\sf Reply-To}, {\sf Server}, {\sf Subject}, {\sf User-Agent}, {\sf Alert-Info}, {\sf Error-Info}, and {\sf Warning} header fields belong to this category. Exactly which of these fields should be removed depends on whether user privacy or service provider privacy needs to be protected. If user privacy is desired, the header fields among them with privacy characteristics ``u'' should be removed. If service provider privacy is desired, the header fields among them with privacy characteristics ``p'' should be removed. If both user and service provider privacy are desired, all these header fields should be removed. For SDP description part (Table~\ref{tab:sdpprivacy}), the {\sf Email}, {\sf Phone}, {\sf Information}, and {\sf URI} fields may be omitted. For RTCP SDES packet fields (Table~\ref{tab:rtpprivacy}), the {\sf Name}, {\sf Location}, {\sf Email}, and {\sf Tool} fields may be omitted.

Anonymization at the user side applies to some of the mandatory fields that could reveal privacy-sensitive information. The exact way to anonymize these fields depends on whether user privacy or service provider privacy needs to be protected. If user privacy is to be protected, the host name or IP address in SIP {\sf Call-ID} and {\sf In-Reply-To} header fields, as well as the host name in RTCP SDES {\sf Cname} field, should be replaced by a suitably long random value; the user display name in SIP {\sf Contact}, {\sf From}, and {\sf To} header fields should be replaced by {\it anonymous}; the SIP URIs that are not used for routing further messages within the same SIP dialog should be anonymized to \verb# sip:anonymous@anonymous.invalid#. As an example, the anonymized form of a {\sf From} header may become \verb# From: "Anonymous" <sip:anonymous@anonymous.invalid>#.

If service provider privacy is to be protected, anoymization should be applied specifically to the service provider related part of the privacy-sensitive information. Therefore, the host name or IP address in SIP {\sf Call-ID} and {\sf In-Reply-To} header fields, as well as the host name in RTCP SDES {\sf Cname} field, should still be replaced by a suitably long random value; the user display name in SIP {\sf Contact}, {\sf From}, and {\sf To} header fields do not need to be worried about; the SIP URIs that are not used for routing further messages within the same SIP dialog should be partially anonymized. Taking the {\sf From} header as an example again, its anonymized form for hiding service provider information may be \verb# From: "Alice" <sip:alice@anonymous.invalid>#. It can be seen that only service provider information is concealed; the original user information is kept intact. One important note about the service provider privacy case is that, as we mentioned in Section~\ref{sec:privacyfield}, the fields we identified here may belong to different types of service providers. The privacy concern may be only for a subset of these service providers. For example, we may be interested in VSP identity hiding, not ISP identity hiding. Unless we have enough information to exactly locate the fields that are only related to the type of service provider we are interested in, we will have to act on all service provider privacy-sensitive fields shown in Section~\ref{sec:privacyfield}.

Note that authentication-related SIP headers need to be treated specially. RFC 3323 discusses these headers as follows: ``Note that authentication mechanisms, including the Digest authentication method described in the SIP specification, are outside the scope of the privacy considerations in this document. Revealing identity through authentication is highly selective, and may not result in the compromise of any private information. Obviously, users that do not wish to reveal their identity to servers that issue authentication challenges MAY elect not to respond to such challenges.''

The second method at the user side to protect privacy is encryption. For SIP messages (including SDP descriptions), RFC 3261 allows the user agent to protect SIP message privacy using S/MIME\cite{rfc3851}. The whole SIP message can be encrypted using the tunneled ``message/sip'' MIME type, or a subset of the SIP message can be encrypted using the ``message/sipfrag'' MIME type. This end-to-end encryption mechanism is often used by the user to convey information to its destination user, without disclosing it to network intermediaries. It is suitable for certain header fields with end-to-end semantic, including {\sf Alert-Info}, {\sf Authentication-Info}, {\sf Error-Info}, {\sf In-Reply-To}, {\sf Organization}, {\sf Reply-To}, {\sf Server}, {\sf Subject}, {\sf User-Agent}, and {\sf Warning}. In some cases, the user also provides an encrypted version for those header fields that already have a clear text version. This is usually because the clear text header field may be modified in the network, but the origin user wants the destination user to receive the unchanged header value. For example, sometimes the SIP {\sf From} header field may be anonymized in the network for privacy reasons. But the origin user wants the destination user to see the original {\sf From} header value so it can properly display ``caller-ID''. This can be solved by inserting an encrypted version of the {\sf From} header when the message is generated.

The encryption method can also be applied to RTP and RTCP. It is defined in RFC 3711 as part of the secure RTP mechanism. %

\subsection{Network Intermediary Privacy Service Functions} \label{sec:nps}

\subsubsection{Network Intermediary-specific Privacy Service Functions}  \label{sec:nispsf}

In both SIP signaling and RTP media, there are privacy-sensitive fields that cannot simply be concealed at the user side without affecting normal operation. For example, the SIP {\sf Contact}, {\sf Via}, {\sf Route}, and {\sf Record-Route} header fields contain URIs used for routing signaling messages within the current SIP dialog, they must be visible to the signaling routing element. The session IP address in SDP {\sf Origin} and {\sf Connection} fields also have to be valid. Finally, although not the focus of our document, some lower level information such as the IP source addresses in RTP and RTCP could reveal identity of the traffic source as well. These information usually can not be manipulated by the user side.

Solving the privacy problem for these fields requires the involvement of network intermediaries. Existing documents suggest the following three methods that the network intermediaries can use for the SIP {\sf Via}, {\sf Route} and {\sf Record-Route} headers. These methods are also applicable to the {\sf Contact} header field.

\begin{description}
\item [Stripping:] RFC 3323 introduces the ``stripping'' method, which removes the corresponding header fields and replaces them with those created by the privacy service. For example, in {\sf Via} stripping, a list of {\sf Via} header fields can be replaced by a single {\sf Via} header field corresponding to the network entity performing the privacy service.
\item [Encryption:] RFC 2543 and Byerly\cite{siphideroute} propose the ``encryption'' method, which asks the privacy service to use a secret key to encrypt the corresponding header fields, with a timestamp and an appropriate checksum. Any network signaling entities beyond the one performing the privacy service will see the encrypted version of these fields. Only the original privacy service entity can decrypt these fields when the message containing these fields comes back.
\item [Caching:] RFC 2543 also mentions a ``caching'' method, which falls somewhere in between the ``stripping'' and ``encryption'' methods. The privacy service keeps a cache of the corresponding header fields and replaces them in the actual message with indexes into the cache. On the reverse path, the privacy service takes the header fields from the cache rather than from the message. It is suggested that the ``encryption'' method is favored over the ``caching'' method because of cache reuse concerns.
\end{description}

Two important conditions need to be satisfied in applying any of the above methods, namely, ``recoverable'' and ``routable''. ``Recoverable'' means the privacy service must make sure that it can restore the values for any of the header fields it has manipulated when further requests or responses within the same SIP dialog arrive. ``Routable'' means that any manipulation of those header fields by a network privacy service needs to make sure further messages within the same SIP dialog can be routed back to the same privacy service for header value recovery. Manipulating header fields in the network and subsequently restoring them necessarily require additional state information to be stored locally. Storage of excessive state information may have significant impact on the scalability of the entity providing privacy service. Among the three methods, the ``encryption'' method imposes the least burden on state management. However, encryption operation may require more processing power and incur longer processing latency.

It should be noted that the ``encryption'' and ``caching'' processing methods are not yet standardized. These methods were originally proposed in RFC 2543 and were then removed in the updated version of RFC 2543, or RFC 3261. We suggest that at least the ``encryption'' method be allowed as part of the SIP privacy framework. In fact, the main counterargument mentioned in RFC 3261 for the ``encryption'' method is ``serious trust issue''. But the trust concern can be alleviated in specific application environments. VoIP peering is one such example where privacy service can be provided by the VoIP peering provider, and VSPs are assumed to trust their peering provider.

In addition to performing privacy service functions on SIP {\sf Via}, {\sf Route} and {\sf Record-Route} headers, the network intermediary privacy service should also make sure that they do not add message fields containing privacy-sensitive information to the messages. Depending on whether user privacy or service provider privacy is concerned, these header fields may include {\sf Alert-Info}, {\sf Call-Info}, {\sf Error-Info}, {\sf Server} and {\sf Organization}.

Network intermediary privacy services for RTP media usually involve a middlebox acting as media relay. It divides the media session into two segments so the original media sender and receiver will not see the opposite side directly. The IP addresses in the SDP {\sf Origin} field and {\sf Connection} fields will need to be modified accordingly.

A common type of network intermediaries that can implement the privacy service is SIP proxy. However, some operations require more functionality than what a SIP proxy can provide. According to Table~\ref{tab:sumhdr}, a SIP proxy is not allowed to modify or delete {\sf Record-Route} and {\sf Via} header fields in SIP requests. Therefore, when these operations are involved, the entity performing network intermediary-specific privacy service functions needs to act as a transparent Back-to-Back User Agent (B2BUA). B2BUA also has to be used when an RTP media relay is needed because in that case the SDP description of the SIP message body must be modified. A SIP proxy is not allowed to modify that part of the SIP message.

\subsubsection{User Side Privacy Service Functions Performed at the Network Intermediary} \label{sec:upsatnps}

If the user side is not able to perform its own privacy service functions, the network privacy service may need to act on the user's behalf. This applies to the removal and anonymization operations discussed in Section~\ref{sec:ups}. Note that the exact operation and the affected fields depend on whether user privacy or service provider privacy needs to be protected.

The network intermediary performing SIP privacy service functions on behalf of the user can be a SIP proxy. But in some cases, it will need to act as a B2BUA. One such case is when any dialog matching field needs to be modified. As mentioned in RFC 3323, modification of the {\sf Call-ID} header field is such an example. Modification of the {\sf From} header field may also cause problem for older SIP implementations. Both {\sf Call-ID} and {\sf From} are header fields that need to be anonymized for privacy. Another case when B2BUA may be needed is when the {\sf Alert-Info}, {\sf Error-Info}, and {\sf Warning} header fields have to be removed. According to Table~\ref{tab:sumhdr}, these fields cannot be deleted by a SIP proxy.
\subsubsection{Indicating Different Network Privacy Service Levels} \label{sec:reqnps}

In general, both the user and the network intermediary may initiate privacy service functions. However, not all user clients and network entities are capable of privacy service functions. There ought to be a mechanism to convey the functions that need to be performed by the network privacy services. RFC 3323 defines a SIP {\sf Privacy} header for that purpose. Five possible values for the {\sf Privacy} header field are specified in RFC 3323. Each of those values is associated with a degree of privacy service as summarized below. {\it User} privacy assumes that the SIP user agent is not capable of privacy service functions, so it requires the network intermediary to perform SIP user side privacy functions protecting user privacy-sensitive information as we discussed in Section~\ref{sec:upsatnps} and Section~\ref{sec:ups}. {\it Header} privacy requires the network privacy service to perform SIP header related privacy functions that can only be done at network intermediaries as specified in Section~\ref{sec:nispsf}. {\it Session} privacy requires the network intermediary to perform RTP media privacy service functions discussed in Section~\ref{sec:nispsf}. {\it None} privacy prohibits any privacy function to be performed. {\it Critical} privacy indicates that the request should be rejected if the corresponding privacy level cannot be accommodated. An additional value for the {\sf Privacy} header is defined in RFC 3325, which is called {\it id} privacy. It means the SIP {\sf P-Asserted-Identity} header should be removed before the message is transfered outside the trusted domain.

We extend the privacy definition of RFC 3323 to cover not only user information, but also service provider information. Consequently, we define a new {\it Service Provider} value for the SIP {\sf Privacy} header to reflect that change. When this privacy value is specified, the network privacy service is required to perform SIP user side privacy service functions protecting service provider information as we discussed in Section~\ref{sec:upsatnps} and Section~\ref{sec:ups}.

There are also privacy service deployment considerations in delivering service provider privacy, especially service provider identity hiding. Usually even if a privacy service is deployed at the border of the network, the identity of the privacy service itself is still exposed to outside the network for routing purposes. Therefore, if a service provider wants to hide its identity, it needs to make sure that the identity of the network privacy service is distinct from its own identity. A common practice would be to use a third party network privacy service. It is also likely that the service provider's privacy service and the third party privacy service are used together. There are at least two benefits for this hybrid scheme. First, this allows the service provider to expose as little information to the outside as possible. The service provider can manage its topology hiding by its own privacy service and use the third party privacy service only for identity hiding. Second, the service provider's privacy service helps to offload part of the third party privacy service's state management burden. This allows the third party privacy service provider to scale better.

The same SIP {\sf Privacy} header can be used to request privacy service from both the service provider and the third party privacy service. We noted that a specific rule in RFC 3323 states the following: ``a {\sf Privacy} header may include each legitimate privacy level value zero or one time. When the privacy functions corresponding to a requested privacy level is performed, the corresponding privacy level value is removed from the {\sf Privacy} header. When the last privacy level value (excluding {\it critical}) is removed, the entire {\sf Privacy} header should be removed.'' In the case where the service provider's own privacy service and the third party privacy service co-exist, the service provider's privacy service should not consider the privacy functions to be finished after its own processing, instead it should preserve the corresponding {\sf Privacy} header values and pass them to the third party privacy service, where further processing will be done and the {\sf Privacy} header value will be removed accordingly.

An alternative to requesting privacy service through the {\sf Privacy} header is to make advanced arrangement with the privacy service provider.

\subsection{Achieving High Level Privacy Requirements} \label{sec:achieve}

The user side and network privacy service functions discussed in Section~\ref{sec:ups} and Section~\ref{sec:nps} form the building blocks to satisfy high level privacy requirements. With a privacy definition covering both the user and the service provider, we identify three common cases of high level privacy requirements and discuss how each of them can be satisfied. Variants of these requirements can be similarly accomplished based on solutions to these cases.

In the first case, a user wants to conceal user privacy-sensitive information both from network intermediaries and from the destination user. This can be achieved by applying the user side removal and anonymization functions for user privacy-sensitive information. If user side privacy service is not available, the network can provide the same functions with the SIP {\it user} privacy value in the {\sf Privacy} header field. These operations provide reasonably good user privacy that is sufficient in most cases, although they do not guarantee that all user information is concealed. For example, these functions do not include anonymizing the SIP URI in the SIP {\sf Contact} header, which is needed for routing signaling messages. If the user wants to ensure strict privacy including those items, it will need to also request the SIP {\it Header} privacy from the network privacy service. Furthermore, if the user side is not able to perform user related privacy functions for RTP/RTCP, it will need to request the SIP {\it Session} privacy from the network to accomplish that.

In the second case, a user wants to conceal user privacy-sensitive information from network intermediaries involved in the communication, but not from its destination user. This case requires the techniques similar to those in the first case above. In addition to that, the user side SIP S/MIME and secure RTP/RTCP privacy function should be applied.

In the third case, a service provider wants to conceal service provider privacy-sensitive information from other service providers involved in the communication. To achieve this, the user side removal and anonymization service for service provider information should be applied. If user side privacy service is not available, the same service should be performed by the network with the SIP {\it service provider} privacy level. The network should also apply SIP {\it header} privacy. In addition, the identity of the network privacy service should be distinct from the identity of the service provider. Finally, the network may need to apply SIP {\it session} privacy in case any RTP/RTCP fields disclose information about the type of the service provider to be protected. Note that if there is no specific user level privacy concerns, the user may apply S/MIME to its SIP messages and allow the destination user to get all user information, without affecting the privacy concerns at the service provider level.
\section{Providing Privacy for Service Providers in VoIP Peering} \label{sec:pps}

\subsection{VoIP Peering Network Architecture} \label{sec:voiparch}

A VoIP peering partner network or VSP network usually consists of a common set of logical functions including Location Function (LF), Policy Function (PF), Signaling Function (SF) and Media Function (MF)~\cite{speermint:arch}. The location function discovers the servers and hosts to be contacted through the help of location services such as DNS or ENUM. The policy function performs authentication and exchange policy parameters used by the signaling function. The signaling function performs the actual routing of SIP signaling messages. The media function may be optional because VoIP signaling and media do not necessarily follow the same path. When the media function is present, it is responsible for operations such as media transcoding and media security enforcement.

\begin{figure}[!thp]
\begin{center}
\subfigure[Composed model]{\label{fig:depmod1.png}
\includegraphics[scale=0.4]{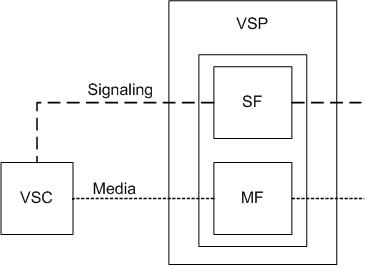}}
\subfigure[Quasi-decomposed model]{\label{fig:depmod2.png}
\includegraphics[scale=0.4]{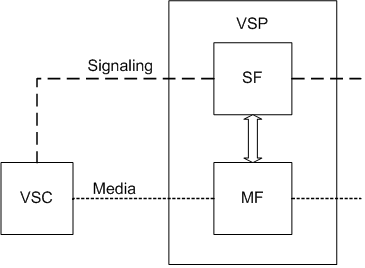}}
\subfigure[Fully-decomposed model]{\label{fig:depmod3.png}
\includegraphics[scale=0.4]{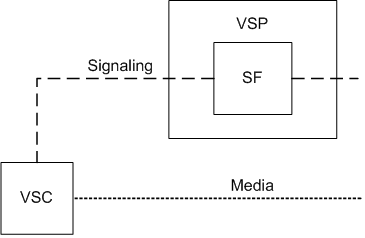}}
\end{center}
\caption{Different VSP deployment models}\label{fig:depmod}
\end{figure}

In a VoIP peering architecture, a VSP has two deployment options based on how the signaling and media relationship is managed, the composed and decomposed model. In the composed model, signaling and media follow the same path. In other words, the SF and MF are implemented in the same peering logical element. The communication between SF and MF in this model is straightforward. The problem with this model is that the combined SF and MF element creates a bottleneck and single point of failure. The decomposed model has distinct signaling and media paths and is more scalable than the composed model. The decomposed model can be further classified into quasi-decomposed and fully-decomposed models. In the quasi-decomposed model, the SF and MF are implemented in separate peering logical elements but usually both of them belong to the same service provider and there is a vertical control interface between the SF and MF. The communication between SF and MF in this model is more complicated than that in the composed model. In the fully-decomposed model, signaling and media are largely independent. There is no direct control between the SF and MF. In fact, the MF may not even need to exist. This model is the most scalable one, but the lack of control between SF and MF may not be desired when operational concerns such as billing and accounting are important. The composed and the two decomposed models of VSP as well as Voice Service Customer (VSC) connecting to it the are illustrated in Figure~\ref{fig:depmod}.

\begin{figure}[!hp]
\begin{center}
\subfigure[Collocated VSP and IAP]{\label{fig:vscp1.png}
\includegraphics[scale=0.5]{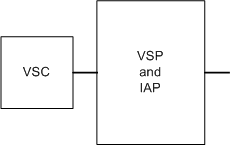}}
\subfigure[Separate VSP and IAP with a single connection]{\label{fig:vscp2.png}
\includegraphics[scale=0.5]{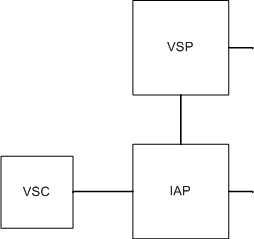}}
\subfigure[Separate VSP and IAP with separate connections]{\label{fig:vscp3.png}
\includegraphics[scale=0.5]{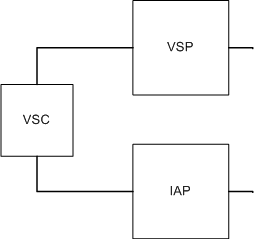}}
\end{center}
\caption{Different VSC and VSP connection methods}\label{fig:vscp}
\end{figure}

Since we are looking at the service provider privacy issue and customers may be associated with more than one service provider, the way customers connect to their service providers is another dimension we need to consider in the VoIP peering architecture. A VSC may be an enterprise network or an end-user. A VSC is usually associated with at least an Internet Access Provider (IAP) and obtains from this IAP data services or voice services or both. Figure~\ref{fig:vscp} shows three common methods that a VSC use to connect to its ISPs. In Figure~\ref{fig:vscp1.png}, the VSC uses a single IAP for both data service and voice service. An example of such type of service providers could be a cable company. In Figure~\ref{fig:vscp2.png} the VSC uses two different service providers for data and voice, but the connection to the VSP is through the IAP. Some residential end users may fall into this category. In Figure~\ref{fig:vscp3.png} the VSC uses two different service providers for data and voice, and the VSC has a separate connection to each of them. This may be the case for an enterprise VSC.

\begin{figure}[!tph]
\centerline{
    \includegraphics[scale=.5]{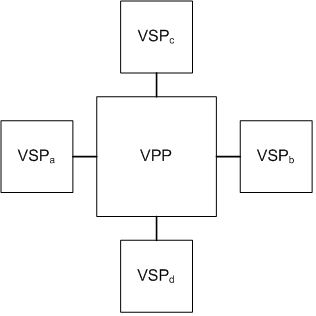}
}
\caption{Peering through a VPP}\label{fig:fed}
\end{figure}

Another dimension of the VoIP peering architecture is whether the VSPs peer directly with each other or they peer through a third party VoIP Peering Provider (VPP). Essentially a VPP is like a VSP. It has the same logical functional components as a VSP and can also be deployed in composed or decomposed models. But a dedicated VPP is a better place to connect multiple VSPs and provide advanced peering services, including a privacy service. A simple illustration of VPP is shown in Figure~\ref{fig:fed}. The use of VPP is particularly relevant in discussing service provider privacy. Although VSP can deploy its own privacy services and achieve topology hiding, identity hiding of a VSP usually requires a third party privacy service, and VPP is the natural place to host this service. Therefore, in the following discussion we focus on a VPP-based VoIP peering architecture. %

\begin{figure}[!tph]
\centerline{
    \includegraphics[scale=.4]{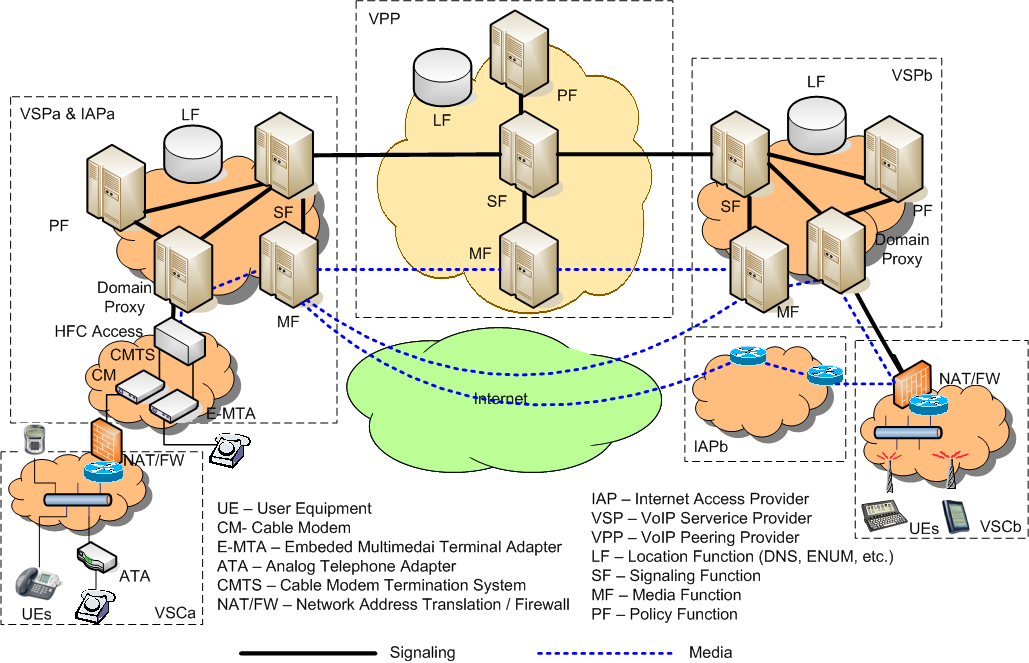}
}
\caption{An example VoIP peering scenario}\label{fig:peeringex}
\end{figure}

A real VoIP peering network scenario is a result of the choices of the VoIP deployment models in Figure~\ref{fig:depmod} and VSC-VSP connection methods in Figure~\ref{fig:vscp}, thus it could be in many different forms. To better understand that, we show an example in Figure~\ref{fig:peeringex}. In this example scenario, VSCa connects to a cable service provider which provides both data and voice services (VSPa and IAPa). VSCb connects to a data service provider (IAPb) and a voice service provider (VSPb) separately. VSCb has a dedicated connection to VSPb. VSPa and VSPb both join the peering service provided by VPP. The signaling paths and media paths in this figure reveal several possible deployment models. At the VSPa side, all media traffic passes through an MF, resulting in the quasi-decomposed model of VSPa. At the VPP and VSPb, the MF element may or may not be used, resulting in either the quasi-decomposed model or the fully-decomposed model.

\subsection{Applying Privacy Mechanisms to VoIP Peering} \label{sec:apppripeer}

\subsubsection{General Rules} \label{sec:apprule}

Assuming we have a generic VPP-based VoIP peering architecture as in Figure~\ref{fig:fed} and the VPP implements the privacy mechanism described in Section~\ref{sec:voipprivacy}, the VSP can either pre-arrange with the VPP for privacy services or dynamically request privacy services from the VPP as follows: If the VSP and IAP are collocated as in Figure~\ref{fig:vscp1.png}, then regardless of the VSP deployment model, both signaling and media privacy need to be protected. The VSP should therefore request {\it header} privacy, {\it service provider} privacy, and {\it session} privacy from the VPP privacy service. If the VSP and IAP are separate as in Figure~\ref{fig:vscp2.png} and Figure~\ref{fig:vscp3.png}, we should further examine the VSP deployment models. If composed or quasi-decomposed model as in Figure~\ref{fig:depmod1.png} and Figure~\ref{fig:depmod2.png} is used, both signaling and media pass through the VSP. Therefore, the VSP should request {\it header} privacy, {\it service provider} privacy, and {\it session} privacy from the VPP privacy service. If fully-decomposed model as in Figure~\ref{fig:depmod3.png} is used, the VSP only handles signaling, so it should request {\it header} privacy and {\it service provider} privacy from the VPP privacy service.

In short, if the VSP only handles signaling, then requesting {\it header} privacy and {\it service provider} privacy should be sufficient. If the VSP handles both signaling and media, then {\it header} privacy, {\it service provider} privacy and {\it session} privacy should all be requested. Note that the requirement to request {\it session} privacy in the latter case could be relaxed if the service provider can afford some risk of its identity being guessed. Specifically, when {\it header} privacy with {\it service provider} privacy are available, an outsider knowing who the IAP is cannot conclude whether that provider is also the VSP for VoIP signaling, unless he already knows the fact that the customer's media and signaling go through the same provider.

Sometimes VSPs also implement their own network privacy functions, usually in their network border elements such as session border controllers. This case is an example of a service provider using a third party privacy service while keeping its own privacy service to perform part of the work, as we already discussed in Section~\ref{sec:reqnps}.

In VoIP peering context, the VSP privacy requirement is usually bi-directional. This is relatively straightforward when the privacy service is provided by a VPP. As long as the VSPs for the caller and callee both follow the rules in Section~\ref{sec:apprule} to pre-arrange privacy service or dynamically ask for appropriate privacy service levels in request and response messages going out of each of their domains, the VPP should be able to provide VSP privacy in both directions.

\subsubsection{VoIP Peering Privacy Example Message Flow}

\begin{figure}[tph]
\centerline{
    \includegraphics[scale=.6]{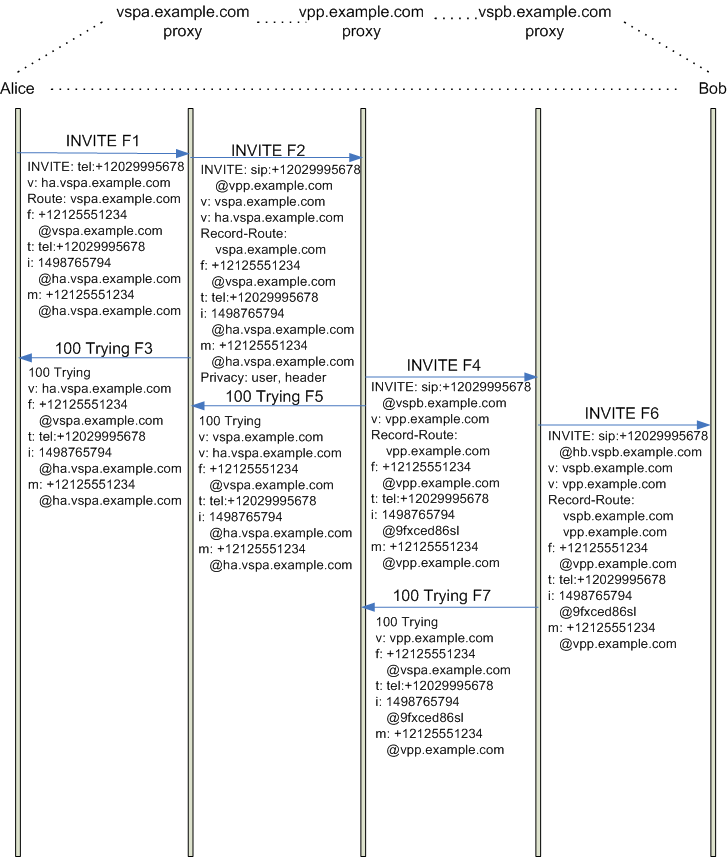}
}
\caption{Example VoIP peering privacy service message flow - part I}\label{fig:egflow1}
\end{figure}

\begin{figure}[tph]
\centerline{
    \includegraphics[scale=.6]{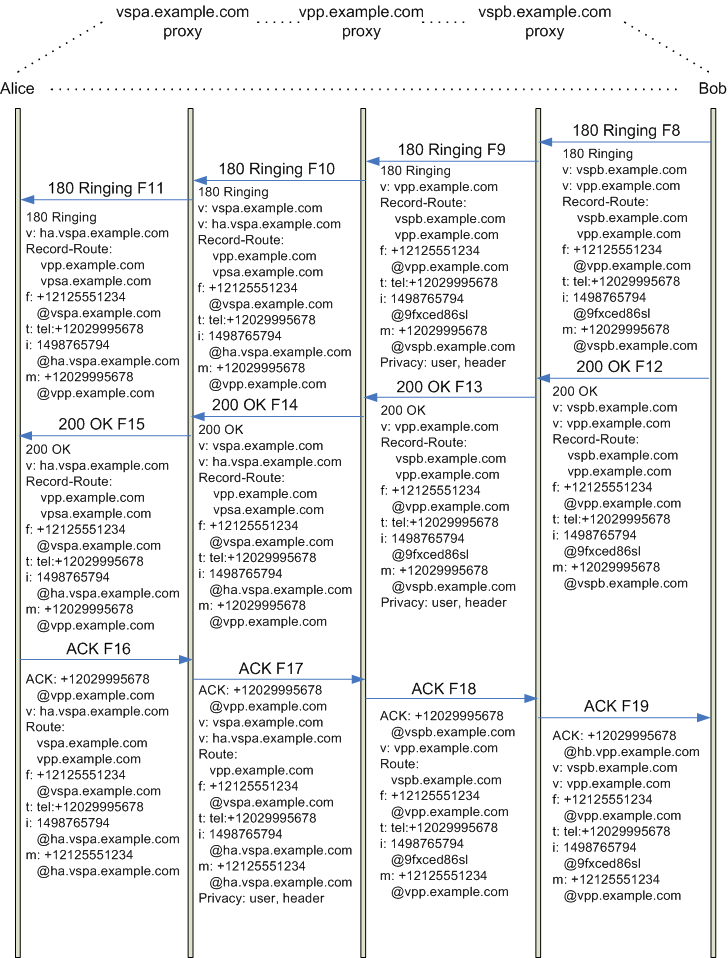}
}
\caption{Example VoIP peering privacy service message flow - part II}\label{fig:egflow2}
\end{figure}

\begin{figure}[tph]
\centerline{
    \includegraphics[scale=.6]{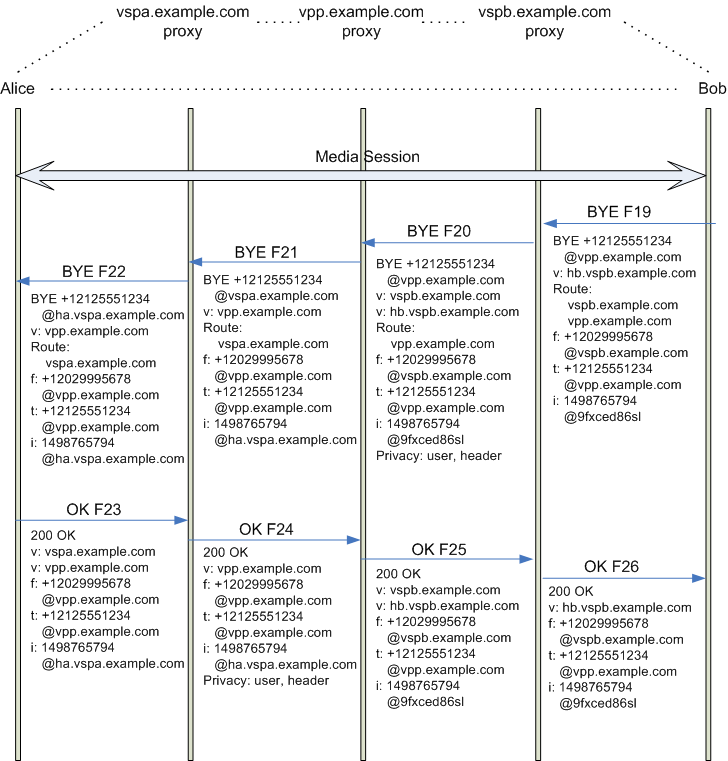}
}
\caption{Example VoIP peering privacy service message flow - part III}\label{fig:egflow3}
\end{figure}

Figure~\ref{fig:egflow1} through Figure~\ref{fig:egflow3} show the detailed message flow of a bi-directional privacy service example in VoIP peering; only the most important header fields are shown. In this example, caller Alice is served by VSPa (vspa.example.com) and callee Bob is served by VSPb (vspb.example.com). VSPa and VSPb peer with each other through VPP (vpp.example.com). VSPa and VSPb both want service provider privacy. Media paths are independent of the signaling paths. Therefore, both VSPa and VSPb request {\it service provider} and {\it header} privacy service levels from VPP. All privacy services are carried out at the VPP. The VPP privacy service implement the privacy mechanism presented in Section~\ref{sec:voipprivacy}, including the ``striping'' method for {\sf Via}, {\sf Route} and {\sf Record-Route} headers. %

\section{Conclusions} \label{sec:conclusion}

Existing work on SIP privacy focuses on privacy-sensitive information for end users. The VoIP peering community needs a privacy mechanism that also cover privacy-sensitive information for VSPs. Specifically, VSPs want to prevent their identity and topology information from being exposed to unintended parties during the VoIP communication. In this report, we summarize and extend related work, and present a unified VoIP privacy mechanism for both user and service providers. We first examined the SIP, SDP, RTP and RTCP message fields that potentially reveal user or service provider information. Then we discussed privacy service functions that can be performed by the user side and the network intermediaries. The user side can apply removal, anonymization or encryption to those privacy-sensitive message fields. The network intermediaries can perform similar functions on behalf of the user side if necessary. In addition, the network intermediaries also need to handle privacy for several routing specific fields, which cannot be done at the user side. The different network privacy functions can be indicated by the SIP {\sf Privacy} header values. We illustrated the use of appropriate {\sf Privacy} header values to achieve user level and service provider level privacy requirements. As an example, we discussed a generic VoIP peering architecture characterized by different options of VoIP deployment and connection models, and show how the VSPs can achieve identity and topology hiding in VoIP peering context using the proposed privacy mechanism.

\section{Acknowledgements}

We thank Fibernet Telecom Group, Inc. for providing funding and equipment for this research.

\bibliographystyle{IEEE}

\bibliography{rfc,i-d-self}

\begin{thebibliography}{10}

\bibitem{rfc3323}
J.~Peterson,
\newblock ``{A Privacy Mechanism for the Session Initiation Protocol (SIP)},''
  RFC 3323 (Proposed Standard), Nov. 2002.

\bibitem{rfc3325}
C.~Jennings, J.~Peterson, and M.~Watson,
\newblock ``{Private Extensions to the Session Initiation Protocol (SIP) for
  Asserted Identity within Trusted Networks},'' RFC 3325 (Informational), Nov.
  2002.

\bibitem{rfc3261}
J.~Rosenberg, H.~Schulzrinne, G.~Camarillo, A.~Johnston, J.~Peterson,
  R.~Sparks, M.~Handley, and E.~Schooler,
\newblock ``{SIP: Session Initiation Protocol},'' RFC 3261 (Proposed Standard),
  June 2002,
\newblock Updated by RFCs 3265, 3853, 4320.

\bibitem{rfc3711}
M.~Baugher, D.~McGrew, M.~Naslund, E.~Carrara, and K.~Norrman,
\newblock ``{The Secure Real-time Transport Protocol (SRTP)},'' RFC 3711
  (Proposed Standard), Mar. 2004.

\bibitem{rfc2543}
M.~Handley, H.~Schulzrinne, E.~Schooler, and J.~Rosenberg,
\newblock ``{SIP: Session Initiation Protocol},'' RFC 2543 (Proposed Standard),
  Mar. 1999,
\newblock Obsoleted by RFCs 3261, 3262, 3263, 3264, 3265.

\bibitem{siphideroute}
{B. Byerly}, {D. Daiker}, and {S. Bhatnagar},
\newblock ``{SIP Record-Route/Route Hiding},''
\newblock Internet draft, Oct. 2000,
\newblock work in progress.

\bibitem{rfc3966}
H.~Schulzrinne,
\newblock ``{The tel URI for Telephone Numbers},'' RFC 3966 (Proposed
  Standard), Dec. 2004.

\bibitem{rfc3550}
H.~Schulzrinne, S.~Casner, R.~Frederick, and V.~Jacobson,
\newblock ``{RTP: A Transport Protocol for Real-Time Applications},'' RFC 3550
  (Standard), July 2003.

\bibitem{rfc3851}
B.~Ramsdell,
\newblock ``{Secure/Multipurpose Internet Mail Extensions (S/MIME) Version 3.1
  Message Specification},'' RFC 3851 (Proposed Standard), July 2004.

\bibitem{speermint:arch}
R.~Penno (Editor),
\newblock ``{SPEERMINT Peering Architecture},''
\newblock Internet draft, Aug. 2006,
\newblock work in progress.

\end{thebibliography}

\end{document}